\documentclass{aastex62}

\received{January 1, 2018}
\revised{January 7, 2018}
\accepted{\today}
\submitjournal{PASP}
\shorttitle{Herschel Disks}
\shortauthors{Tanner et al.}

\begin{document}

\title{Herschel Observations of Disks Around Late-type Stars}

\correspondingauthor{Angelle Tanner}
\email{angelle.tanner@gmail.com}

\author[0000-0002-0786-7307]{Angelle Tanner}
\affil{Mississippi State University, Department of Physics \& Astronomy, Hilbun Hall, Starkville, MS, 39762, USA}

\author[0000-0002-8864-1667]{Peter Plavchan}
\affil{Department of Physics and Astronomy, George Mason University 4400 University Avenue, Fairfax, VA 22030, USA}

\author[0000-0000-0000-0000]{Geoff Bryden}
\affil{Jet Propulsion Laboratory, California Institute of Technology, 4800 Oak Grove Drive, Pasadena, CA 91125, USA}

\author[0000-0001-6831-7547]{Grant Kennedy}
\affil{Department of Physics, University of Warwick, Gibbet Hill Road, Coventry, CV4 7AL, UK}

\author[0000-0000-0000-0000]{Luca Matr{\'a}}
\affil{Harvard-Smithsonian Center for Astrophysics, 60 Garden Street, Cambridge, MA 02138, USA}

\author[0000-0000-0000-0000]{Patrick Cronin-Coltsmann}
\affil{Department of Physics, University of Warwick, Gibbet Hill Road, Coventry, CV4 7AL, UK}

\author[0000-0001-8014-0270]{Patrick Lowrance}
\affil{IPAC, MS 314-5, California Institute of Technology, 1200 E. California Blvd., Pasadena, CA 91125, USA}

\author[0000-0002-9061-2865]{Todd Henry}
\affil{RECONS Institute, Chambersburg, Pennsylvania, 17201, USA}

\author[0000-0000-0000-0000]{Basmah Riaz}
\affil{Universitats-Sternwarte Munchen, Ludwig-Maximilians-Universitat, Scheinerstr. 1, 81679 Munchen, Germany}

\author[0000-0002-8916-1972]{John E. Gizis}
\affil{Department of Physics and Astronomy, University of Delaware, Newark, DE 19716, USA}

\author[0000-0003-1645-8596]{Adric Riedel}
\affil{California Institute of Technology, 770 S. Wilson Avenue, Pasadena, CA 91125, USA}

\author[0000-0000-0000-0000]{Elodie Choquet}
\affil{Aix Marseille Univ., CNRS CNES, LAM, P\^ole de l'\'Etoile Site de Ch\^ateau-Gombert
38, rue Fr\'ed\'eric Joliot-Curie 13388 Marseille Cedex 13 France}

\begin{abstract}

A set of twenty late-type (K5-M5) stars were observed with the Herschel Space Observatory at 100 and 160 microns with the goal of searching for far-infrared excesses indicative of the presence of circumstellar disks. Out of this sample, four stars (TYC 7443-1102-1, TYC 9340-437-1, GJ 784 and GJ 707) have infrared excesses above their stellar photospheres at either 100 or 160 $\micron$ or both. At 100 microns TYC 9340-437-1 is spatially resolved with a shape that suggests it is surrounded by a face-on disk. The 100 $\micron$ excess flux associated with GJ 707 is marginal at around 3$\sigma$. The excess flux associated with GJ 784 is most likely due to a background galaxy as the dust radius estimated from the spectral energy fit implies that any associated dust disk should have been resolved in the Herschel images but is not. TYC 7443-1102-1 has been observed with ALMA which resolves the emission at its location into two distinct sources making the Herschel excess most likely also due to a background galaxy.  It is worth noting that this star is in the 23 Myr old $\beta$ Pic association. With a disk luminosity on the order of 10$^{-3}$ L$_*$, this system is an ideal follow-up target for high-contrast imaging and ALMA.

\end{abstract}


\section{Introduction}

 Low-mass stars such as late-K and all M-dwarfs are the most populous stars in our local universe, and yet we are just now determining how often they form planetary systems (Dressing \& Charbonneau 2015; Dressing et al. 2019). These stars play a critical role in understanding planet formation and evolution as the two competing planet formation paradigms, core accretion (Pollack et al. 1996) and gravitational instability (Boss 1997), predict very different planet frequencies for the smallest stars. Theoretical investigations regarding the formation of planets around M dwarfs make it clear that the predicted number of Jupiter and terrestrial mass planets and the timescale for planet formation significantly depend on the formation scenario and initial conditions (Pascucci et al. 2011). Core accretion models predict relatively few giant planets (1\%; Kennedy \& Kenyon 2008), while gravitational instability models (e.g. Boss 2006) predict giant-planet frequencies similar to those of Solar-type stars, perhaps 10\% or more. Another planet formation scenario, which involves pebble accretion, has been successfully used to recreate the compact planetary system around the late-M, Trappist-1 star with an outer disk radius of 200 AU (Ormel et al. 2017; Schoonenberg et al. 2019). All of these planet formation mechanisms rely on the protostellar disk environment as an initial condition to planet formation. Identifying and characterizing protoplanetary disks across a wide range of ages and stellar environment is key to connecting these planetary formation theories to the types of planetary systems we see today (Roberge et al. 2012; Ertel et al. 2020). 
 
  Both transit and radial velocity surveys have estimated giant planet occurrence rates which are less that what is found for FGK stars with a 2-3 times smaller occurrence rate, for instance, for 0.3-3M$_J$ planets with period less than 100 days (Bonfils et al. 2013; Cumming et al. 2008). The occurrence rate is different for smaller planets (1-4 R$_{\earth}$) with Dressing \& Charbonneau (2015) finding more than one planet (2.5$\pm$0.2) expected to be present around low-mass stars in orbits less than 50 days. Given the difference in the outcome of planetary architectures from formation models as well as an observed difference in planet populations between giant and rocky planets, knowing as much as possible about the properties of disks around low-mass stars over an ample age range provides insight into the dominant initial conditions for planet formation. From an observational standpoint, it is advantageous to characterize the amount of dust around these stars to determine to what degree exo-zodical emission will inhibit the direct detection of planets near the star.

Relative to earlier-type stars, the frequency and properties of M-dwarf debris disks are observationally poorly constrained. We do know that more than half of low-mass T Tauri stars as young as 1 Myr possess primordial disks (Meyer et al. 1997; Haisch et al. 2001), and primordial disks are known to persist for longer around M dwarfs than earlier-type stars (Carpenter et al. 2006). Debris disks around mature M dwarfs, however, are rare with more disks found so far around G- and A-type stars (Su et al., 2006; Meyer et al. 2006; Thureau et al. 2014; Sibthorpe et al. 2018). The sample of M dwarfs older than 10 Myr with clear disk detections long-ward of $\lambda\sim$70 $\micron$ is growing (i.e. Plavchan et al. 2005; Gautier et al. 2007; Lestrade et al. 2006; Smith et al. 2006; Zuckerman et al. 2011; Kalas et al. 2005; Rodriguez et al. 2014; Silverberg et al. 2016; Binks \& Jefferies 2017; Kennedy et al. 2018; Bayo et al. 2019). Selection effects, such as luminosity contrast and lower effective dust temperatures relative to earlier-type stars, have not completely explained the potential lack of $>$10 Myr M dwarfs with observable circumstellar material (Liu et al. 2004; Plavchan et al. 2005); however, sensitivity could play a role to some degree (Luppe et al. 2019). Currently, we do know that circumstellar disks around young stars evolve and dissipate, but the physical mechanisms responsible and the timescale for disk dispersion are not yet observationally constrained. On the other hand, stellar age could be playing a role as, on average, most field late-type stars are older thus allowing for additional collisional evolution within the disk which would lead to a lower luminosity for the same initial disk mass (i.e. Marino et al. 2019).  Here, we present the results of a small Herschel imaging survey of young or nearby M dwarfs. In \S 2 we discuss the selection of our stellar sample, in \S 3 we discuss the Herschel observations and data analysis, and in \S 4 we review the results of the survey.

\section{Sample}

Our late-type dwarf sample consists of 20 stars falling into two categories: 1) late-type dwarfs with known mid-infrared excesses and 2) late-type dwarfs that show indications of being young ($<$ 300 Myr) and are nearby ($<$ 40 pc) and, therefore, have a good probability of having circumstellar disks which are detectable with Herschel. Table~\ref{targettable} contains the list of stars in this program. Because our project was submitted to the last GO Call for the Herschel satellite, we chose a sample of specific science targets compared to a flux-limited or volume-limited survey. The stars in our sample have a distance range of 4-50 pc, a range in K$_s$ magnitudes of 5 to 8.5 and encompass stellar spectral types of K5 to M5V. The young stars in the sample have been characterized as such due to their rapid rotation rates and large X-ray fluxes (Plavchan et al. 2009).  

\section{Observations and Analysis}

For this survey, we used the Herschel Photoconductor Array Camera and Spectrometer (PACS) photometer (Pilbratt et al. 2010; Poglitsch et al. 2010) in small scan-map mode to image debris disk candidate sources using the 100 and 160 $\micron$ band configuration. We selected the 100/160 micron configuration of PACS to maximize the signal to noise of our observations and hence the positional accuracy (using the 0.6$\times$FWHM/SNR estimator). We are also taking advantage of the stability in the shape and full width half max (FWHM) of the point spread function (PSF) at these longer wavelengths compared to 70 $\micron$ (2-4\% at 100 $\micron$ versus 10\% at 70 $\micron$, Kennedy et al. 2012). The dominant error in the astrometry is the pointing accuracy of the telescope. Because the cold dust we expected to detect with Herschel should have spectral energy distributions that peak between 100 and 200 $\micron$ the signal to noise improvement gained by observing at 100 $\micron$ outweighs the marginal ($\sim$40\%) reduction in FWHM gained by working at 70 $\micron$. Each observation consists of two scans performed in different directions to provide a cross-scan. The medium scan rate is used (20 $\arcsec$/sec), with a scan leg length of 3 arcminutes, a cross-scan step of 4$\arcsec$, and 10 scan legs per map repeat. The resulting pixel scale of the mosaicked images is 1$\arcsec$/pixel for the 100 $\micron$ images and 2$\arcsec$/pixel for the 160 $\micron$ images. These parameters are taken from the PACS mini scan-map mode release note (PICC-ME-TN-036, v2.0, 12th Nov. 2010). Each map was repeated 6 times to reach the desired noise level of 2.75 and 5.24 mJy at 100 and 160 $\micron$, respectively. The depth of these observations is comfortably above the confusion noise in these bands. 

\subsection{Photometry}

The Herschel photometry is extracted using apertures of 5$\arcsec$ and 10$\arcsec$ in radius for 100 and 160 $\micron$, respectively, and a sky annulus from 20 to 40$\arcsec$ in radius. These radii are the similar to those used by larger Herschel programs for unresolved sources including the DUst around NEarby Stars (DUNES) survey which used radii of 5 and 8$\arcsec$ at these two wavelengths (Eiroa et al. 2013).The photometric uncertainty is calculated by directly sampling the variation of the background flux away from the target star. Using the same aperture size and aperture correction as for the target star photometry, the background brightness is measured throughout the surrounding sky annulus. In order to account for the correlated noise between neighboring pixels, the uncertainty in the photometry values is estimated by convolving the sky annulus with the photometric aperture size and then estimating its RMS. The signal to noise ratio of the detected flux ($\chi$) is then used to determine whether that flux is significantly greater than the stellar photosphere. The significance of an infrared excess is quantified via $\chi$ = $(F_{obs} - F_{*}) / \sqrt{\sigma_{obs}^2 + \sigma_{*}^2}$ where F and $\sigma$ are the flux and errors of the stellar photosphere and observed flux in that band pass. Table~\ref{phottable} provides the aperture photometry estimated from our Herschel images, $\chi$ values and excess flux above the photosphere, $F_{obs}/F_{*}$, in each pass band. In Table~\ref{phottable} we also provide estimates of the ratio of the dust luminosity to the stellar luminosity, $L_d$/$L_*$, for those stars with both confirmed infrared excesses and non-detections using the formulation given in Beichman et al. (2006). For the stars with non-detections, the $L_d$/$L_*$ values are upper limits. 

\subsection{Spectral Energy Distribution (SED) Fitting}

The optical and near-infrared photometry for each star is extracted from multiple sources in the literature including TYCHO (H$\o$g et al. 2000), DENIS (Epchtein et al. 2007) and 2MASS (Cutri et al. 2003). The optical stellar fluxes are fitted using a grid of K and M dwarf PHOENIX NextGen spectra (Allard \& Hauschildt 1995) which depend on stellar temperature while assuming Solar metallicity. The SED used to model the Herschel photometry is a simple blackbody function given there are at most only two far-infrared data points for our excess stars. The best fit to the complete SED is achieved with the amoeba minimization algorithm by minimizing the $\chi^2$ value comparing the fit of the model to the observed fluxes. The complete SED of the star and the dust is those components added together and fit to the optical and infrared fluxes. To provide a more physical fit to the data, we utilized the assumption of radiative equilibrium for the dust grains which relates the luminosity of the star (via its temperature and radius) to the temperature and distance of the dust grains. Therefore, the free parameters in our fit are the stellar radius and temperature and dust distance and temperature. The first guesses for the stellar radius and temperature are taken from the Gaia catalog (Gaia Collaboration et al. 2018). The range in which these values are allowed to vary during the fit come from the range of values given in this catalog. The values of the free parameters which produce the best fits to the photometry are provided in Table~\ref{sedtable}.

A systematic calibration uncertainty (2.75\% and 4.15\% for PACS100 and PACS160, respectivelyis also included in the SED fitting  (PICC-ME-TN-037 Herschel release note; Balog et al. 2014).  The uncertainties on the fitted parameters are estimated by varying each parameter until the reduced chi square value changed by one and holding the remaining free parameters constant at their best fitting values. We also provide a lower limit to the dust mass derived from Jura et al. (1995, equation 5) using our estimate of $L_d$/$L_*$ and assuming spherical grains with a density of 3.5 g/cm$^3$, and radius of 1.0 $\micron$. In the case of an object which is not detected at either 100 or 160 $\micron$, a dust luminosity limit is estimated by setting the peak of a blackbody curve to occur at 160 $\micron$, adjusting the dust radius until the combined stellar and dust SED passes through the 160 $\micron$ upper limit and adding up the total flux within that blackbody curve. 

\section{Results}

Out of the 20 stars that were observed with Herschel, four (TYC 7443-1102-1, TYC 9340-437-1, GJ 784 and GJ 707) have infrared excesses near or above 3$\sigma$ (see Table~\ref{phottable}). TYC 7443-1102-1, TYC 9340-437-1, and GJ 784 have a significant flux detections at both 100 and 160 $\micron$ while GJ 707 is only detected at 100 $\micron$. Below we discuss each detection in detail as well as the significance of the non-detection of the nearby young star, AP Col. 

\subsection{TYC 9340-437-1}

TYC 9340-437-1 is a $\beta$ Pic Moving Group member at an age of 23$\pm$3 Myr (Gaia Collaboration et al. 2018; Mamajek \& Bell 2014; Torres et al. 2006) and a distance of 36.66$\pm$0.03 pc (Gaia Collaboration et al. 2018). With fluxes of 88.9$\pm$1.7 and 98.6$\pm$6.5 mJy at 100 and 160 $\micron$, respectively, this star has significant flux excesses of 51 and 15$\sigma$, respectively, at 100 and 160 $\micron$. In addition, the disk is spatially resolved at 100 $\micron$ (see Figure~\ref{hersch2}). The FWHM of the dust emission is 23$\pm$1$\arcsec$ compared to the Herschel PSF FWHM of 6.77$\arcsec$ at this wavelength (Poglitsch et al 2010; PACS Observers Manual). An image of the subtraction of the Herschel PSF from the image of TYC 9340-437-1 is also shown in Figure~\ref{hersch2}. The remaining flux appears to be symmetric at this spatial resolution. 

The convolution of a simple ring model with the PSF model is subtracted from the observed Herschel image and is also shown in Figure~\ref{hersch2}. The ring model is offset by 1.4$\arcsec$ to match the difference in the nominal and observed target positions (see Table 2). The dust disk is assumed to be radially concentrated into a thin ring with ($\Delta$R/R) of 0.1, a Gaussian radial profile, and an inclination is 7$\pm$19 degrees consistent with a face-on disk. Because of its face-on orientation, the position angle of the disk is not well defined. We use an MCMC procedure (emcee; Foreman-Mackey 2013) to find both the best fit parameters and the range of values that are consistent with the observations. The successful fit of a ring model produces a radial spatial extent of the dust of 2.6$\pm$0.2$\arcsec$ corresponding to a dust radius of 96$\pm$7.4 AU. The absence of significant residuals from this model subtraction in combination with the non-detection of any extended emission in HST/NICMOS observations suggests that the orientation of the circumstellar material is consistent with a face-on disk. When we consider the luminosity of the star (0.19 L$_{\sun}$) and the measured radius of the dust disk, this data lies above the power-law fit but within the scatter of data points in Figure 1 of Matr{\`a} et al. (2018) which addresses a potential correlation between these two observables. If we use the best fit line to the Matra et al (2018) data, we'd expect this disk to have a radius of $\sim$50 AU. Our single new data point does not significantly impact the conclusions from this study that this correlation indicates that there are preferred radii for the formation of planetesimals. The age of this star and the large dust radius could, however, indicate the presence of planetesimal stirring in the absence of a planetary system (Krivov \& Booth 2018).

Our SED fit to the photometry for this system results in a dust temperature and radius of 37$\pm$10 K and 24$\pm$6 AU, respectively. If we force the equilibrium dust radius to the 96 AU estimated from the resolved image, then we can fit all of the photometry again with the same dust temperature but now with a 1/$\lambda$ emissivity power law for the dust grains and a grain radius of 1.25 $\micron$. With only two photometry points to characterize the dust, we are hesitant to fit the data with the more complex dust models used in similar studies (e.g. Morales et al. 2016, Dodson-Robinson et al. 2016).  To estimate the probability that this detection could be due to a background galaxy, we have consulted Table 2 in Sibthorpe et al. (2013) which provides an integration of the galaxy luminosity function at 100 $\micron$. From this table we estimate that this source has a probability of $\sim$1.5$\times$10$^{-5}$ of being a background source. Because this star was pre-selected based on the presence of its mid-IR excess, it is more likely than a randomly chosen star to be contaminated by a far-IR background source. Nevertheless, we conclude, based on the brightness of the observed far-infrared flux and the fact that the emission is spatially resolved, that the likelihood of such contamination is small.

%stellar luminosity: 0.18544 Lsun, -0.74 -> 50 AU from Matra

\subsection{TYC 7443-1102-1}

TYC 7443-1102-1 is also a $\beta$ Pic Moving Group member at a distance of 51.24$\pm$0.12 pc and an age of 23$\pm$3 Myr (Gaia Collaboration et al. 2018; Mamajek \& Bell 2014). The PACS images show a solid detection right at the stellar position. The fluxes are 10.2$\pm$1.4 mJy at 100 $\micron$ and 17.2$\pm$4.5 mJy at 160 $\micron$. This is $\sim$23 times the photosphere at 100 $\micron$. Fits to the SED of the star produce a dust temperature and radius of 27$\pm$4 K and 37$\pm$12 AU, respectively. This inner dust radius determined from the SED fit is consistent with the point source seen in the Herschel image which corresponds to an upper limit on the disk radius of $<$ 200 AU as determined at 100 $\micron$. 

TYC-7443-1102-1 was observed with ALMA in band 7 (0.87 mm, 345 GHz) on 22 August 2018 under project 2017.1.01583.S (PI: G. Kennedy). The observations used baselines ranging from 15.1 to 483.9 m from 48 antennae with an average precipitable water vapor of $\sim$0.76 mm. The total on-source observing duration was 14.62 minutes. J2056-4714 and J1924-2914 were used for pointing, bandpass and flux calibration. J1924-2914 was observed between individual target scans for time-varying atmospheric calibrations. The spectral setup comprised four windows centered on 347.883, 335.883, 333.987 and 345.987 GHz with a width 2 GHz and 128 channels for all but the last window which has a width 1.875 GHz and 3840 channels.

The raw data was calibrated with the provided ALMA pipeline script in CASA version 5.1.2-4 (McMullin et al. 2007). All images were generated with the CLEAN algorithm in CASA. Continuum imaging was carried out using natural weighting and self-calibration was not attempted. This gives a synthesized beam with a position angle of 83.6$^\circ$ and semi-major and semi-minor axes of 0.43$\arcsec$ and 0.48$\arcsec$ respectively. The corresponding image is shown in Figure \ref{ALMA}, with a measured rms of 40 $\mu$Jy beam$^{-1}$. At the wavelength of observation the star would have a flux density of $4.5 \mu$Jy and is not detected. Two distinct sub-mm sources are clearly detected, neither are centered at the Gaia DR2 proper-motion adjusted location of the star. The two sources are 1.4$\arcsec$ and 0.9$\arcsec$ distant from the stellar location and have flux densities of 1.3 mJy and 0.5 mJy respectively. There is also a smaller peak to the west of the stellar location separated by 0.17$\arcsec$ and with a flux density of 0.1 mJy.

The ALMA absolute pointing accuracy for this observation is $\sim$30 mas and the error on the Gaia stellar location is sub-milliarcsecond. Therefore, these mm-wave sources are most likely not associated with the star and constitute background galaxies. For a putative debris disk to be detected with PACS but not with ALMA, the spectral slope of the dust emission would need to have $\gamma \lesssim -1$, steeper than is seen for well-characterised cases (e.g. G$\acute{a}$sp$\acute{a}$r et al. 2012; Macgregor et al. 2016). Larger surveys (that are less precise) find $\gamma$ values in the range of $-0.5$ to $-1$ (Holland et al. 2017). Thus a scenario where the PACS detection is of a circumstellar disk that is then not detected by ALMA is improbable. Therefore the Herschel excess most likely originated from these contaminating sources and not from a circumstellar disk.

\subsection{GJ 707}

GJ 707 (HIP 89211, HD166348) is a M0V star at a distance of 13.20$\pm$0.01 pc (Gaia Collaboration et al. 2018). This star has significant excess emission in just the 100 $\micron$ band. The observed flux at this wavelength is 12.7$\pm$2.7 mJy which gives it a 3.3$\sigma$ excess above the stellar photosphere. This star is also in the Moro-Martin et al. (2015) paper with a flux of 11.3$\pm$2.8 mJy. While our flux measurements are consistent, the excess detection at 100 $\micron$ is marginal at best. The star is not detected in the publicly available Spectral and Photometric Imaging Receiver (SPIRE) data from the Disc Emission via a Bias Free Reconnaissance in the Infrared and Sub- millimetr (DEBRIS) survey. Our blackbody fit to the Herschel photometry produces a dust temperature and radius of 65$\pm$6 K and 6.6$\pm$2.1 AU, respectively. The hotter temperature and corresponding smaller equilibrium dust radius than those estimated from the other excess stars in our sample is most likely a result of the fit to the single 100 $\micron$ photometry point. The Herschel image is unresolved at 100 $\micron$ allowing us to place an upper limit of $<$7$\arcsec$ or $\sim$92 AU on the diameter of its purported disk assuming the above stated stellar distance. 

\subsection{GJ 784}

GJ 784 (HIP 99701, HD 191849) is an M0V star at a distance of 6.161$\pm$0.002 pc (Gaia Collaboration et al. 2018). With our data analysis, this star has a $>$3$\sigma$ excess emission at both 100 and 160 $\micron$. At 100 $\micron$ we estimate a flux of 24.5$\pm$1.4 mJy which corresponds to a 10 $\sigma$ detection above the stellar photosphere which is 10.6 mJy at this wavelength. The Herschel DEBRIS survey also included this source in its survey although with a shorter integration time (1686 seconds vs. 721 seconds). Moro-Martin et al. (2015) report a 100 $\micron$ flux of 19.42$\pm$2.95 mJy for this star. At this flux the detection is then at 2.95$\sigma$ significance above the stellar photosphere. Additionally, the peak of the emission seen in our Herschel image is offset with respect to the nominal Herschel position by 2-3 $\arcsec$ - within the pointing error for the Herschel telescope (S{\'a}nchez-Portal et al. 2014). At the location of this star there are SPIRE data also from Herschel in which there is significant emission at 250, 350, and 500 $\micron$, respectively. When we fit an SED to the Herschel data the resulting inner radius and temperature for the dust are 27$\pm$2 K and 25$\pm$5 AU, respectively (see Figure~\ref{seds1}). The Herschel image is unresolved at 100 $\micron$ allowing us to place an upper limit of $<$7$\arcsec$ or $\sim$43 AU on the diameter of the disk assuming the Gaia distance to the star. 

\subsection{AP Col}

AP Col has been categorized as one of the closest pre-main sequence stars through its lithium equivalent width, photometric variability, astrometry and rotational velocity (Riedel et al. 2011) which is why we included it in our sample. It has an age of $\sim$40 Myr and is at a distance of 8.4$\pm$0.07 pc based on RECONS parallax measurements (Reidel et al. 2011, 2014). Despite its relative young age, this star showed no significant dust emission at either 100 or 160 $\micron$. The photometry values for this source are 3.7$\pm$1.6 and 1.0$\pm$5 mJy at 100 and 160 $\micron$, respectively, corresponding to a 1.39 and 0.2 $\sigma$ significance at these two wavelengths. We can use the lack of excess emission at these wavelengths to place an upper limit of L$_d$/L$_*$ $<$ 5$\times$10$^{-5}$ on any dust emission if we use the 100 $\micron$ photometry value and L$_d$/L$_*$ $<$ 6$\times$10$^{-6}$ if we use the 160 $\micron$ photometry value. 

%100 micron - 4.851209779291254e-05
%160 micron - 6.292336346109294e-06

\section{Discussion and Conclusions}

The dust temperatures for TYC 7443-1102-1 and TYC 9340-437-1 are unusually cold for Solar-type stars but less so for dust around early-type M dwarfs (see Figure~\ref{seds1}, Binks \& Jefferies 2017). Despite a small sample size of 20 stars we report significant, resolved excess dust emission associated with the young late-type dwarf, TYC 9340-437-1, at 100 and 160 $\micron$ indicative of the presence of a face-on circumstellar disk. This debris disk has a very cool dust temperature ($\sim$40 K) making it analogous to cold Kuiper belts. This system joins a small collection of additional low-mass stars in the $\beta$ Pic association for which debris disks have been identified via infrared excess (i.e. Binks \& Jefferies 2017). We also report the confirmation of excess emission around two nearby M dwarfs (GJ 784 and GJ 707); however, these detections are either marginally significant or most likely due to background extended emission. Our survey adds one new detection and three non-detections to the sample of 18 members of the $\beta$ Pic moving group observed by Riviere-Marichalar et al. (2014) as part of the GASPS survey (Dent et al. 2013). Their sample includes five late type stars with one (HIP 11437) exhibiting a significant excess dust emission. Therefore, three out of nine late-type stars in the Beta Pic moving group have debris disks detected in the far-infrared with Herschel. HIP 11437 is an K6V star with a disk that is not spatially resolved in the Herschel data with an estimated dust radius of 8.3$\pm$1.7 AU and an L$_d$/L$_*$ of 10$^{-3}$. This means that while its infrared excess is comparable to TYC 9340-437-1, the disk is not as large. Our survey also included AP Col which has no measurable infrared excess emission at 70 $\micron$ having detected the stellar photosphere and an upper limit measured at 160 $\micron$. 

Late-type stars are still an underrepresented group in large, long wavelength stellar debris disk surveys but are quickly catching up to Solar-type stellar surveys. The Herschel DEBRIS survey (Moro-Martin et al. 2015; Sibthorpe et al. 2018) is volume limited for M and K dwarfs out to 8.6 and 15.6 pc, respectively. Their M dwarf survey has yet to be published and contains 117 nearby late-type stars (Phillips et al. 2010). While studies have shown there is no significant correlation between the presence of both debris disks and extrasolar planets (Wyatt et al. 2012; Moro-Martin et al. 2015; Meshkat et al. 2017), this relationship has not been fully explored in terms of stellar mass and age but is slowly improving in sample size and variety (Wyatt et al. 2012; Marshall et al. 2014; Matthews et al. 2014). A recent analogous survey of late-type stars with known planets found two new debris disks in a sample of 21 systems (Kennedy et al. 2018). Our survey did not detect any excesses around our nine field late-type dwarfs, however, this small sample size and a lack of published limits on the presence of planets in these systems prevents any meaningful conclusions on a planet-disk correlation. 

This study provides an additional direct imaging target in TYC 9340-437-1 for ALMA. This star may also be an interesting target for HST STIS observations in search of the disk in scattered light if its large fractional luminosity, L$_d$/L$_*$, $\approx$10$^{-3}$  is enough to counteract the fact that it appears to be a face-on disk in the Herschel images. Further work will be necessary to model the disk in scattered light. This disk can be added to the small collection of circumstellar disks around late-type stars which have been resolved in thermal emission. These include the famous young edge-on disk AU Mic (Matthews et al. 2015), and the debris disk around the planetary system, GJ 581 (Lestrade et al. 2012). Both of these disks were also resolved in Herschel PACS images. There is also the ALMA survey by Long et al. (2018), which resolved disks around a sample of ten late-type stars. By growing the collection of resolved circumstellar disks, we can compare and contrast the morphological properties and grain properties of the disks as a function of age, inclination, and host star – and now also among the late-type stars.
 
\begin{acknowledgements}

We thank the anonymous referee for their insightful comments.

Support for this work, as part of the NASA Herschel Science Center data analysis funding program, was provided by NASA through a contract issued by the Jet Propulsion Laboratory, California Institute of Technology to Mississippi State University. PACS has been developed by a consortium of institutes led by MPE (Germany) and including UVIE (Austria); KU Leuven, CSL, IMEC (Belgium); CEA, LAM (France); MPIA (Germany); INAF-IFSI/OAA/OAP/OAT, LENS, SISSA (Italy); IAC (Spain). This development has been supported by the funding agencies BMVIT (Austria), ESA-PRODEX (Belgium), CEA/CNES (France), DLR (Germany), ASI/INAF (Italy), and CICYT/MCYT (Spain).

This publication makes use of data products from the Wide-field Infrared Survey Explorer, which is a joint project of the University of California, Los Angeles, and the Jet Propulsion Laboratory/California Institute of Technology, funded by the National Aeronautics and Space Administration.  

This paper makes use of the following ALMA data: ADS/JAO.ALMA\#2011.0.01234.S. ALMA is a partnership of ESO (representing its member states), NSF (USA) and NINS (Japan), together with NRC (Canada), MOST and ASIAA (Taiwan), and KASI (Republic of Korea), in cooperation with the Republic of Chile. The Joint ALMA Observatory is operated by ESO, AUI/NRAO and NAOJ.

   This work has made use of data from the European Space Agency (ESA) mission {\it Gaia} (\url{https://www.cosmos.esa.int/gaia}), processed by the {\it Gaia} Data Processing and Analysis Consortium (DPAC, \url{https://www.cosmos.esa.int/web/gaia/dpac/consortium}). Funding for the DPAC has been provided by national institutions, in particular the institutions participating in the {\it Gaia} Multilateral Agreement.
   
   GMK is supported by the Royal Society as a Royal Society University Research Fellow. 
   
   This research made use of Astropy, a community-developed core Python package for Astronomy (Astropy Collaboration 2013)
   as well as Matplotlib (Hunter 2007).

\end{acknowledgements}

\newpage

\begin{deluxetable}{lrccccl}		
\rotate
\footnotesize
\tablecaption{Herschel Sample \label{targettable}}														
\tablehead{														
\colhead{Star Name} & \colhead{Coordinates} & \colhead{Distance$^a$} & \colhead{$V$}   & \colhead{$K_{s}$}     & \colhead{SpTy}  & \colhead{Association$^{b,c}$}   \\																
                                  & \colhead{RA/Dec (2000)} &   \colhead{parsecs} & \colhead{mag} & \colhead{mag}     &              																
                                   }																
\startdata																
SCR0103-5515	 &	01 03 35.5 $-$55 15 55.6	&	45.17	&	9.83	&	7.21	&	M5V	&	Tuc-Hor	\\
2MASS J02543316-5108313AB	&	02 54 33.1 $-$51 08 31.2	&	...	&	12.21	&	7.79	&	 M1.5V	&	Tuc-Hor\\
GJ 141	&	03 24 59.7 $-$05 21 49.5	&	15.63	&	7.84	&	5.12	&	K4V	&	Field	\\
LP776-025	&	04 52 24.4 $-$16 49 21.0	&	14.06	&	11.61	&	6.89	&	M3.0V	&	AB Dor	\\
AP Col	&	06 04 52.2 $-$34 33 36.1	&	8.39	&	12.96	&	6.87	&	M4.5V	&	Argus	\\
GJ 234 A	&	06 29 23.4 $-$02 48 50.3	&	4.13	&	11.07	&	5.49	&	M4.5	&	Field	\\
2MASS J06434532-6424396	&	06 43 45.3 $-$64 24 39.6 	&	...	&	13.00	&	8.37	&	M3V	&	Field	\\
GJ 412 A	&	11 05 28.6	 +43 31 36.4	&	4.85	&	8.78	&	4.77	&	M2.0	V&	Field	\\
GJ 514	&	13 29 59.8 +10 22 37.8	&	7.69	&	9.03	&	5.04	&	M1.0V	&	Field	\\
GJ 617 A	&	16 16 42.7 +67 14 19.8	&	10.64	&	8.60	&	4.95	&	M1V	&	Field	\\
GJ 638	&	16 45 06.4 +33 30 33.2	&	9.80	&	8.11	&	4.71	&	K5V	&	Field	\\
GJ 707	&	18 12 21.4 $-$43 26 41.5	&	13.16	&	8.38	&	5.24	&	M0V	&	Field	\\
GJ 725 A	&	18 42 46.7 +59 37 49.5	&	3.57	&	8.91	&	4.43	&	M3V	&	Field	\\
TYC 7443-1102-1	&	19 56 04.4 $-$32 07 37.7	&	...	&	11.80	&	7.85	&	K9IV	&	$\beta$ Pic	\\
GJ 784	&	20 13 53.4 $-$45 09 50.5	&	6.21	&	7.97	&	4.28	&	M0V	&	Field	\\
L755-019	&	20 28 43.4 $-$11 28 29.1	&	18.20	&	12.47	&	7.50	&	M3.5V	&	Argus	\\
AT Mic	&	20 41 51.2 $-$32 26 06.8	&	10.75	&	10.34	&	4.94	&	M4V	&	$\beta$ Pic	\\
HD 358623	&	20 56 02.8 $-$17 10 53.9	&	...	&	10.62	&	7.04	&	K6V	&	$\beta$ Pic	\\
Gl 821	&	21 09 17.4 $-$13 18 09.0	&	12.20	&	10.86	&	6.91	&	M1.0V	&	Field	\\
TYC 9340-437-1	&	22 42 48.9 $-$71 42 21.3	&	37.00	&	10.60	&	6.89	&	K7V	&	$\beta$ Pic	\\
\enddata		
\tablenotetext{a}{All distances are from from Hipparcos unless otherwise stated in the text (van Leeuwen 2007).}
\tablenotetext{b}{The ages corresponding to the stated young associations or individual stars include $\sim$40 Myr for Tuc-Hor (Kraus et al. 2014), 12-50 Myr for AP Col in Argus (Riedel et al. 2011), 23 Myr for $\beta$ Pic (Torres et al. 2006; Mamajek and Bell 2014) and $\sim$50 Myr for AB Dor (Luhman et al. 2005).}
\tablenotetext{c}{All of the field stars in our sample exhibit excess emission in Spitzer observations as do AT Mic, HD 358623 and TYC 9340-437-1 which are all in the $\beta$ Pic association.}
																										
\end{deluxetable}		
																					
\newpage

\begin{deluxetable}{lc|rcc|rcc|rc}	
\rotate
\footnotesize
\tablecaption{Herschel Photometry \label{phottable}}																										
\tablehead{	
\colhead{Star name} & \colhead{Offset$^a$}  &  \multicolumn{3}{c}{100 \micron}                                                                                 & \multicolumn{3}{c}{160 \micron}                                                                                          & \colhead{L$_d$/L$_*$}  & \colhead{Other Herschel} \\
                                  & \colhead{arcseconds} & \colhead{$F_{\nu}$ (mJy)} & \colhead{$F_{\nu}/F_{\star}$} & \colhead{$\chi$ $^b$} & \colhead{$F_{\nu}$ (mJy)}         & \colhead{$F_{\nu}/F_{\star}$} & \colhead{$\chi$ $^b$} & \colhead{10$^{-5}$}       & \colhead{Notes}}																									
\startdata	    
SCR0103-5515	&	2.0	&	3.6	$\pm$1.7	&	36.24	&	2.06	&	4.2	$\pm$	4.9	&	108.24	&	0.8	&	$<$94	&		\\			
2MASS J02543316-5108313AB	&	0.0	&	-0.3$\pm$	1.6	&	-0.77	&	-0.43	&	-9.9	$\pm$	4.6	&	-65.25	&	-2.2	&	$<$1.5	&		\\			
GJ 141	&	0.0	&	5.1$\pm$	2.4	&	1.40	&	0.61	&	11.6	$\pm$	4.7	&	8.20	&	2.1	&	$<$1.1	&	DEBRIS	\\			
LP776-25	&	4.2	&	3.4	$\pm$	1.8	&	2.74	&	1.20	&	-1.1	$\pm$	4.2	&	-2.27	&	-0.3	&	$<$6.8	&		\\			
AP Col	&	0.0	&	3.7	$\pm$	1.6	&	2.51	&	1.39	&	1.0	$\pm$	5	&	1.74	&	0.2	&	$<$6.2	&		\\			
GJ 234 A	&	2.6	&	7.1	$\pm$	1.4	&	1.70	&	2.09	&	6.7	$\pm$	3.5	&	4.20	&	1.4	&	$<$3.8	&		        \\			
2MASSJ06434532-6424396	&	0.0	&	-2.5	$\pm$	1.6	&	-6.96	&	-1.79	&	14.8	$\pm$	6.2	&	105.56	&	1.65	&	$<$13	&		\\			
GJ 412 A	&	4.2	&	6.3	$\pm$	2.5	&	0.90	&	-0.28	&	4.7	$\pm$	6.2	&	1.80	&	0.3	&	$<$1.6	&	DEBRIS	\\			
GJ 514	&	2.9	&	7.7	$\pm$	2.5	&	1.50	&	1.03	&	-0.5	$\pm$	6.6	&	-0.30	&	-0.4	&	$<$2.6	&	DEBRIS	\\			
GJ 617 A	&	1.5	&	7.2	$\pm$	2.1	&	1.50	&	1.14	&	-4.4	$\pm$	4.3	&	-2.40	&	-1.5	&	$<$2.6	&	DEBRIS	\\			
GJ 638	&	3.2	&	5.6	$\pm$	2.3	&	1.00	&	0.00	&	-2.2	$\pm$	4.7	&	-1.00	&	-0.9	&	$<$0.9	&	DEBRIS	\\			
{\bf GJ 707}	&	2.3	&	12.7	$\pm$	2.7	&	3.40	&	3.32	&	9.1	$\pm$	7	&	6.30	&	1.1	&	$<$4.3	&	DEBRIS	\\			
GJ 725 A	&	0.1	&	9.4	$\pm$	1.5	&	0.60	&	-4.18	&	10.2	$\pm$	4.6	&	1.70	&	0.9	&	$<$1.1	&	DEBRIS	\\			
{\bf TYC7443-1102-1}	&	0.8	&	10.2	$\pm$	1.4	&	23.41	&	6.97	&	17.2	$\pm$	4.5	&	101.03	&	3.9	&	35.4	&		\\			
{\bf GJ 784}	&	2.8	&	23.6	$\pm$	1.2$^c$	&	2.35	&	10.04	&	30.4	$\pm$	4.5	&	7.45	&	5.9	&	5.8	&	DEBRIS	\\			
L755-19	&	4.2	&	3.4	$\pm$	1.8	&	4.81	&	1.50	&	-1.1	$\pm$	4.2	&	-3.98	&	-0.3	&	$<$12	&		\\			
AT Mic	&	2.0	&	10.1	$\pm$	1.2	&	1.50	&	2.81	&	5.0	$\pm$	2.5	&	1.90	&	1	&	$<$3.3	&	GASPS	\\			
HD 358623	&	0.0	&	-1.5	$\pm$	1.9	&	-1.52	&	-1.31	&	0.8	$\pm$	5.5	&	2.07	&	0.2	&	$<$2.8	&		\\			
GJ 821	&	1.4	&	-0.6	$\pm$	1.7	&	-0.56	&	-0.99	&	-8.7	$\pm$	6.8	&	-20.68	&	-1.3	&	$<$0.96	&		\\			
{\bf TYC9340-437-1}	&	1.4	&	88.9	$\pm$	1.7	&	102.95	&	51.79	&	98.6	$\pm$	6.5	&	292.32	&	15.1	&	121.8 &		\\			
\enddata																										
\tablenotetext{a}{Offset between the observed and nominal target positions. Values of 0.0 indicate no peak detected within 5$\arcsec$.}													
\tablenotetext{b}{Significance of any excess (values above 3$\sigma$ shown in boldface).}
\tablenotetext{c}{This flux is the average of the photometry extracted from both our 100 $\micron$ image and that collected as part of the 
DEBRIS survey (Moro-Martin et al. 2015). }																										
\end{deluxetable}	

\newpage

\begin{deluxetable}{lcccccc}	
\tablecaption{SED Model Parameters \label{sedtable}}
\tablehead{\colhead{Star} &\colhead{T$_{\rm eff}$}&\colhead{R$_*$}       &\colhead{T$_d$}  &\colhead{R$_d$} & \colhead{Dust Mass Lower Limit}  \\
                                        &\colhead{Kelvin}    &\colhead{R$_{\sun}$}&\colhead{Kelvin}  &\colhead{ AU   } & \colhead{M$_\earth$/10$^{-6}$} 
}	
\startdata			
GJ 707                      & 4261$\pm$15 & 0.66$\pm$0.02 & 65$\pm$6 &  6.6$\pm$2.1 & $>$0.31 \\
GJ 784                      & 3918$\pm$10 & 0.53$\pm$0.04 & 27$\pm$2 &  25$\pm$5     & $>$2.46 \\
TYC 7443-1102-1   & 3731$\pm$35 & 0.86$\pm$0.06 & 27$\pm$4 &  37$\pm$12    & $>$90 \\
TYC 9340-437-1     & 3949$\pm$45 & 0.91$\pm$0.05 & 37$\pm$10 & 24$\pm$6     & $>$129 \\
\enddata	
\end{deluxetable}			

%%%%%%%%%%%%%%%%%%%%%%%%%%%%%
\begin{figure}[ht]
\plotone{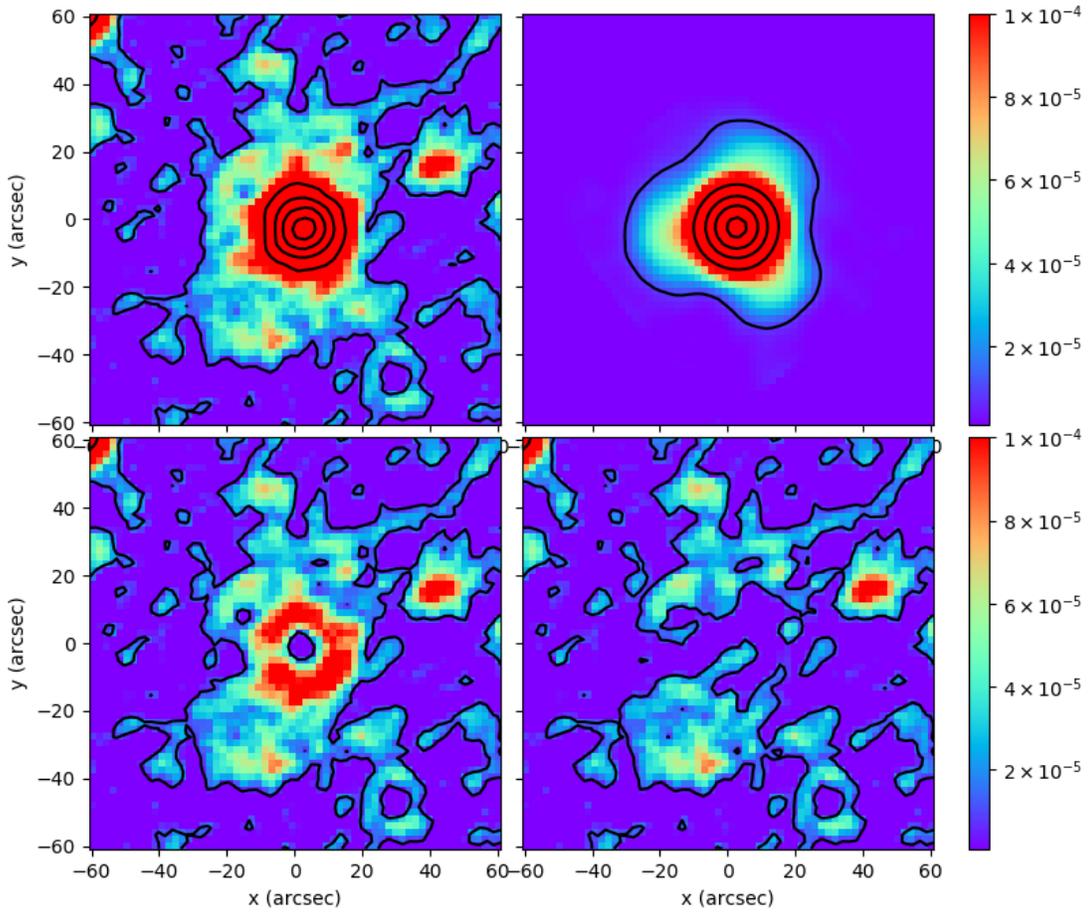}
\figcaption{Herschel images at 100 microns showing 120$\times$120 $\arcsec$ regions around TYC 9340-437-1. The
frames show the the raw data (upper left), the data after subtracting a point-spread
function (lower left), a ring model convolved with the instrument PSF (upper right), and
the residuals when subtracting the model from the data (lower right). All images have the
same linear-stretch scale and each is oriented with North up and East to the left. Note the
ring-like residuals after PSF subtraction, evidence a face-on extended disk. The color bar on the right and the contour levels correspond to the pixel values in Jy. The contour levels are in increments of 10\% of the peak pixel value.   
 \label{hersch1}}
\end{figure}
%%%%%%%%%%%%%%%%%%%%%%%%%%%
%%%%%%%%%%%%%%%%%%%%%%%%%%%%%
\begin{figure}[ht]
\plotone{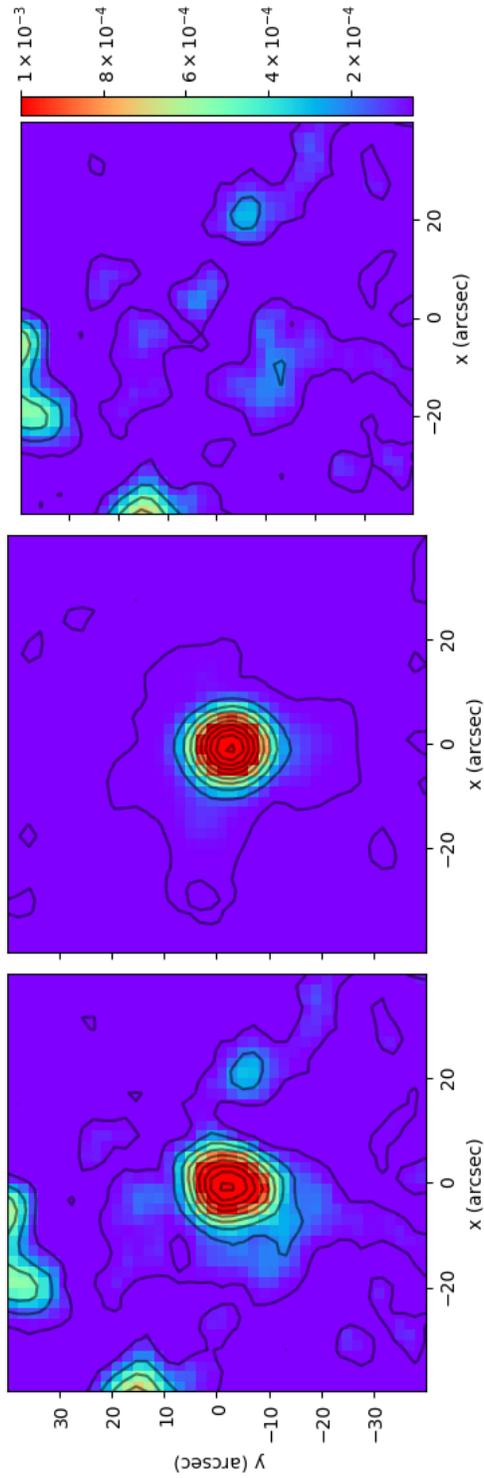}
\figcaption{Herschel images at 160 microns showing 80$\times$80 $\arcsec$  regions around TYC 9340-437-1. The
frames show the the raw data (left), a point-spread function fit to the data (center), and
the data after subtracting the point-spread function (right). All images have the same
linear-stretch scale and each is oriented with North up and East to the left. The color bar on the right and the contour levels correspond to the pixel values in Jy. The contour levels are in increments of 10\% of the peak pixel value.   
 \label{hersch2}}
\end{figure}
%%%%%%%%%%%%%%%%%%%%%%%%%%%
%%%%%%%%%%%%%%%%%%%%%%%%%%%%%
\begin{figure}[ht]
\plottwo{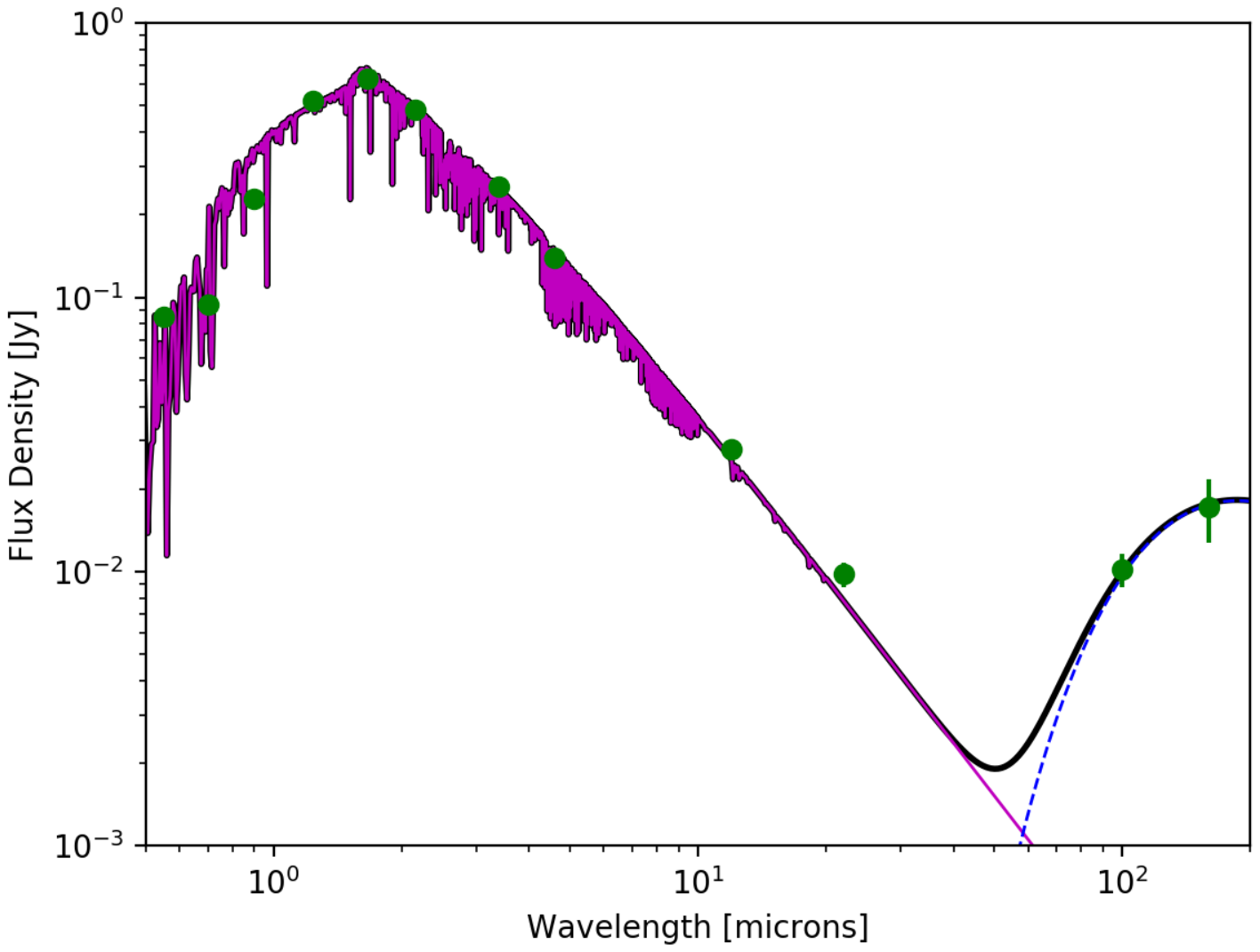}{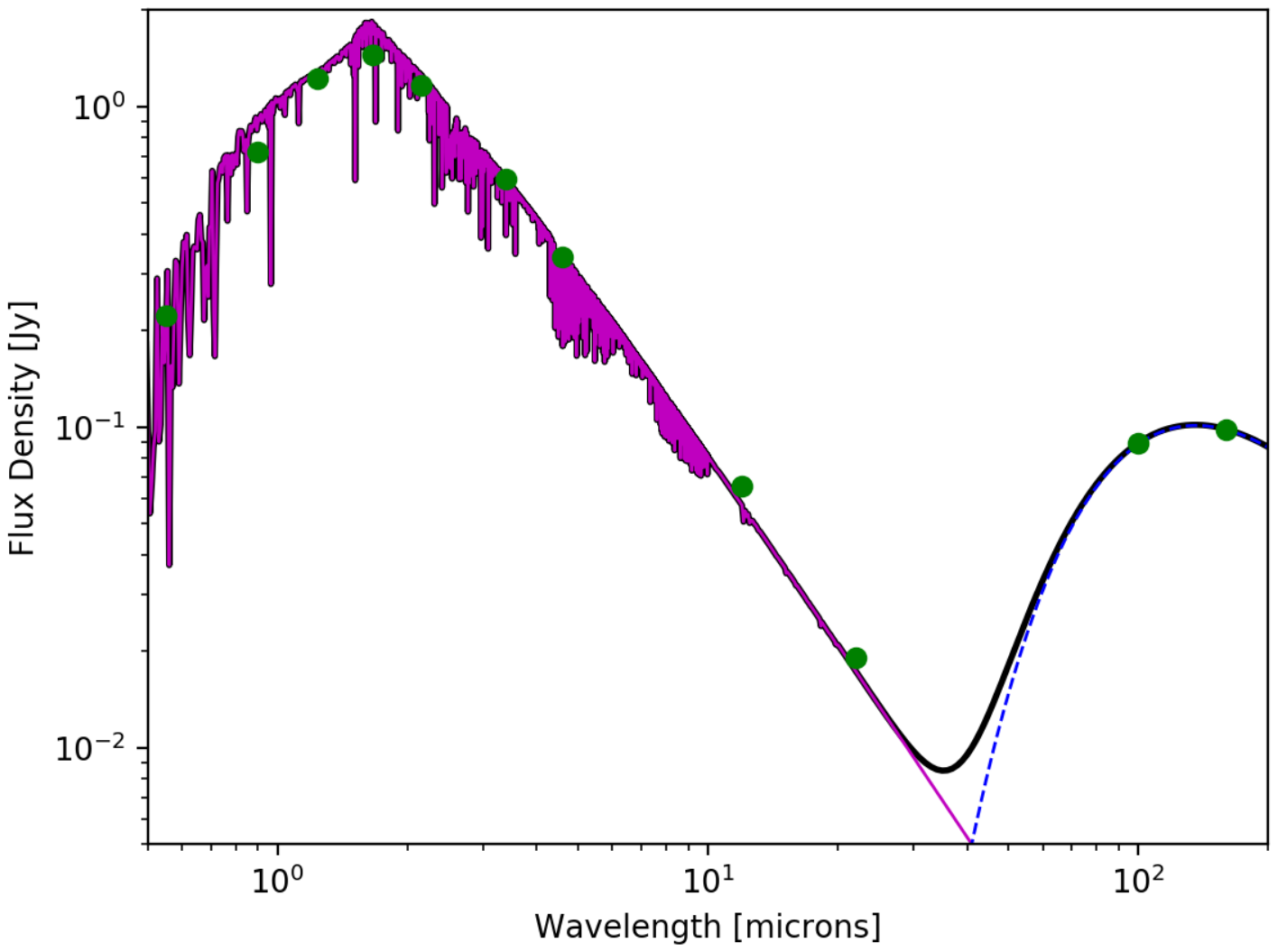}
\figcaption{{\bf Left -} Plot of the photometry and best fitting SED for TYC 7443-1102-1. {\bf Right -} Plot of the photometry and best fitting SED for TYC 9340-437-1.
The purple (solid) line including the shaded region represents the emission from the star, while the blue (dotted) line represents the best blackbody fit to the Herschel photometry. The black (thick solid) line is the sum of the stellar and dust components.  \label{seds1} }
\end{figure}
%%%%%%%%%%%%%%%%%%%%%%%%%%%
%%%%%%%%%%%%%%%%%%%%%%%%%%%%%
\begin{figure}[ht]
\plottwo{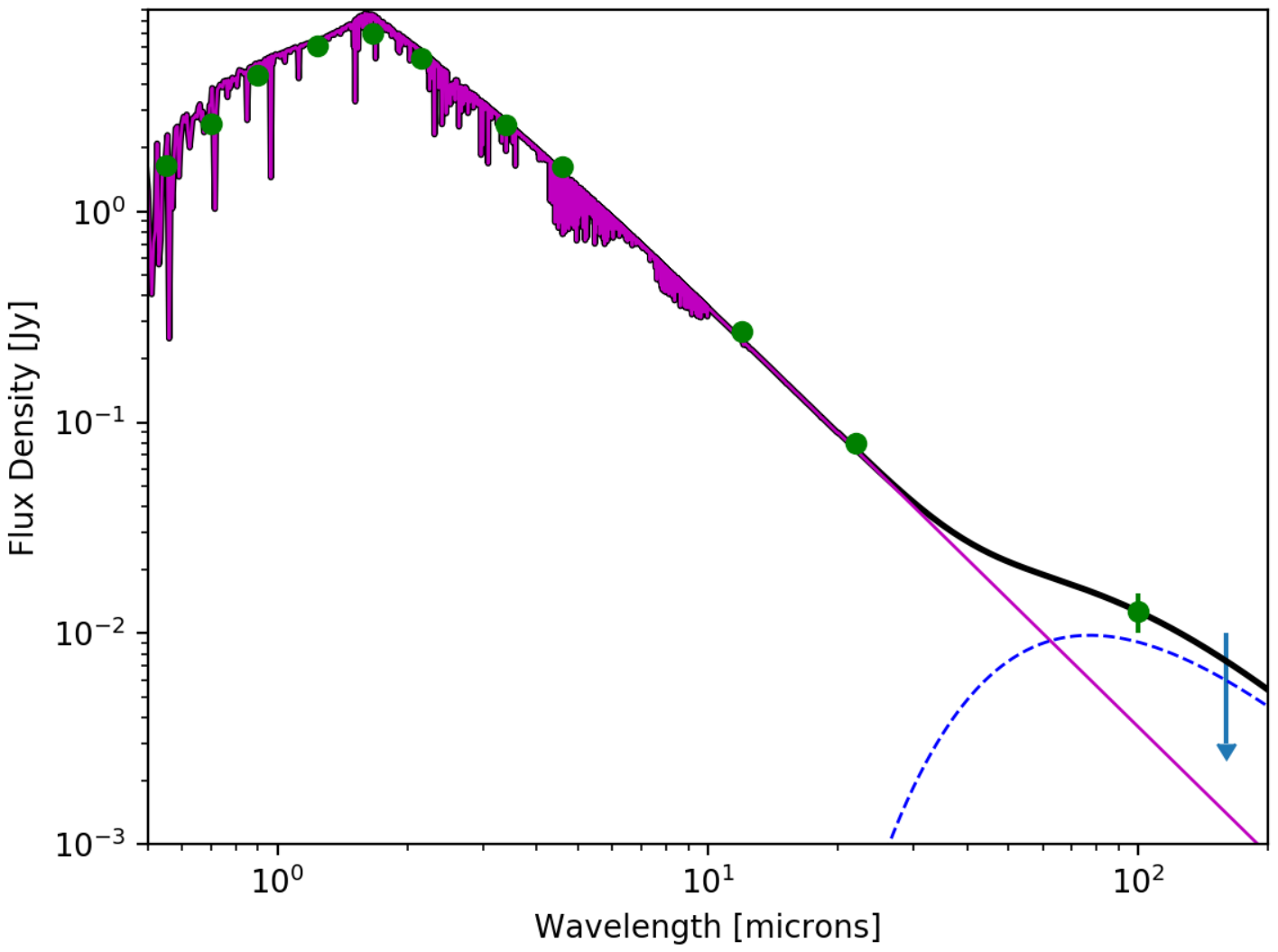}{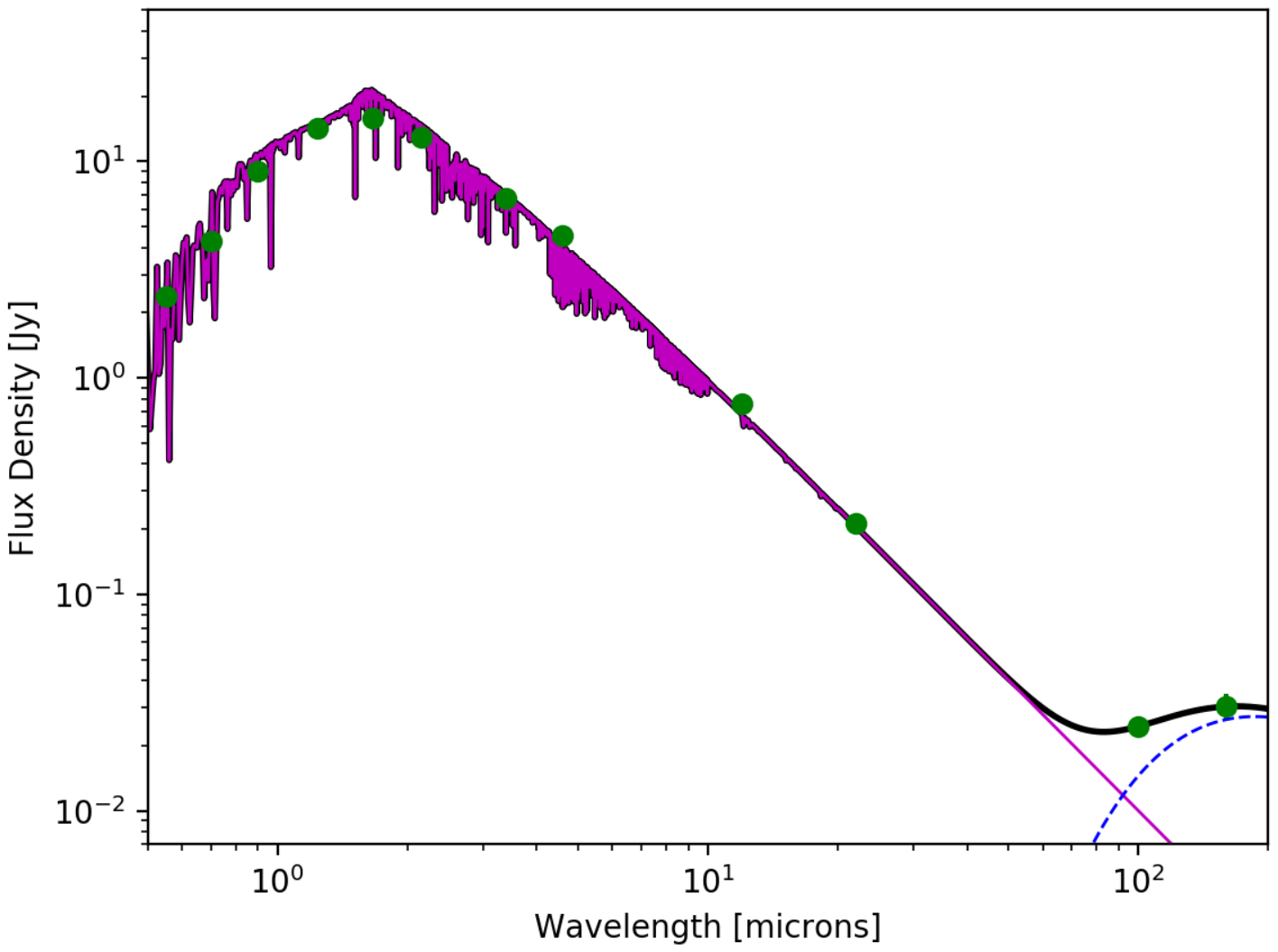}
\figcaption{{\bf Left -} Plot of the photometry and best fitting SED for GJ 707. {\bf Right -} Plot of the photometry and best fitting SED for GJ 784. The purple (solid) line including the shaded region represents the emission from the star, while the blue (dotted) line represents the best blackbody fit to the Herschel photometry. The black (thick solid) line is the sum of the stellar and dust components.  \label{seds2} }
\end{figure}
%%%%%%%%%%%%%%%%%%%%%%%%%%%
%%%%%%%%%%%%%%%%%%%%%%%%%%%%%
\begin{figure}[ht]
\plotone{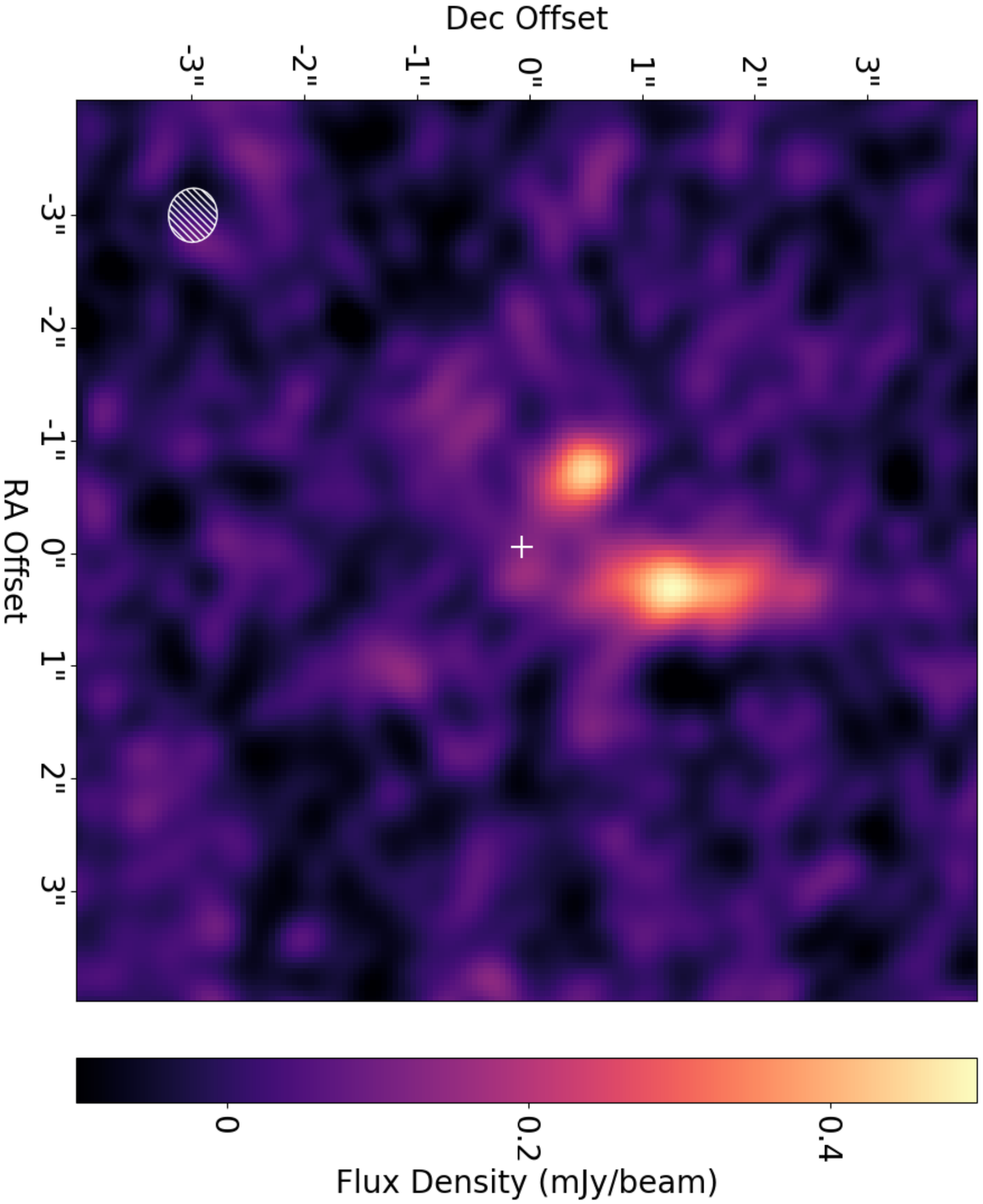}
\figcaption{\label{ALMA} Natural-weighted ALMA image of the sub-mm sources in the vicinity of TYC-7443-1102-1. The ellipse in the lower left corner shows the beam size of 0.43$\times$0.48 $\arcsec$. No emission centered on the stellar location, marked with a $+$, is detected. Two distinct sources well separated from the stellar location are apparent. North is up and east is to the left in this image.}
\end{figure}
%%%%%%%%%%%%%%%%%%%%%%%%%%%

%Krivov 2013 (http://adsabs.harvard.edu/abs/2013ApJ...772...32K)
%is an example of a Herschel debris disk paper that uses
%Herschel galaxy counts to estimate the frequency of contamination.
%They get the frequency of galaxies from Berta 2011
%(http://adsabs.harvard.edu/abs/2011A%26A...532A..49B).
%They discuss how they do this on page 8 (search for 'Berta').
%In particular they use Berta 2011 Figure 7 to estimate the
%background galaxy frequency.
%Looking at that figure, they get
%"we have estimated the density number of sources in the 6–13 mJy
%range to be about 5500 sources per square degree (1.53 sources per
%square arcmin)"
%So the likelihood of finding something ~10 mJy within the
%Herschel FWHM (6 arcsecond radius) is
%1.53 per square arcmin =
%1.53 / 3600 per square arcsec =
%1.53 / 3600 * (pi 6^2) per beam =
%just 5% likely to happen.


\begin{references}
\reference{} Allard, F., \& Hauschildt, P.~H.\ 1995, ApJ, 445, 433
\reference{} Astropy Collaboration, Robitaille, T.~P., Tollerud, E.~J., et al.\ 2013, A\&A, 558, A33
\reference{} Balog, Z., M{\"u}ller, T., Nielbock, M., et al.\ 2014, Experimental Astronomy, 37, 129
\reference{} Bayo, A., et al., 2019, MNRAS, 486, 5552 
\reference{} Beichman, C.~A., Bryden, G., Stapelfeldt, K.~R., et al.\ 2006, ApJ, 652, 1674 
\reference{} Berta, S., et al. \ 2011, A\&A, 532, 49
\reference{} Binks, A.~S., \& Jeffries, R.~D.\ 2017, MNRAS, 469, 579 
\reference{} Bonfils X., et al. \ 2013, A\&A, 549, A109
\reference{} Boss, A. P. 1997, Science, 276, 1836
\reference{} Boss, A.~P.\ 2006, ApJ, 643, 501
\reference{} Burgasser, A.~J., Kirkpatrick, J.~D., Reid, I.~N., et al.\ 2003, ApJ, 586, 512
\reference{} Carpenter, J.~M., Mamajek, E.~E., Hillenbrand, L.~A., \& Meyer, M.~R.\ 2006, ApJL, 651, L49 
\reference{} Cumming A., et al. \ 2008, PASP, 120, 531
\reference{} Cutri, R.~M., Skrutskie, M.~F., van Dyk, S., et al.\ 2003, VizieR Online Data Catalog, 2246, 
\reference{} Dent, W.~R.~F., Thi, W.~F., Kamp, I., et al.\ 2013, PASP, 125, 477
\reference{} Dodson-Robinson, S.~E., Su, K.~Y.~L., Bryden, G., et al.\ 2016, ApJ, 833, 183
\reference{} Dressing, C.~D., et al. \ 2019, AJ, 158, 87
\reference{} Dressing, C.~D., \& Charbonneau, D.\ 2015, ApJ, 807, 45
\reference{} Eiroa, C., Marshall, J.~P., Mora, A., et al.\ 2013, A\&A, 555, A11
\reference{} Endl, M., Cochran, W.~D., K{\"u}rster, M., et al.\ 2006, ApJ, 649, 436
\reference{} Epchtein, N., Deul, E., Derriere, S., et al.\ 1999, A\&A, 349, 236 
\reference{} Ertel, S., Defr{\`e}re, D., Hinz, P., et al.\ 2020, AJ, 159, 177
\reference{} Foreman-Mackey, D., Hogg, D.~W., Lang, D., et al.\ 2013, PASP, 125, 306
\reference{} G{\'a}sp{\'a}r A., Psaltis D., Rieke G. H., Özel F., 2012, ApJ, 754, 74
\reference{} Gautier, T.~N., III, Rieke, G.~H., Stansberry, J., et al.\ 2007, ApJ, 667, 527 
\reference{} Haisch, K.~E., Jr., Lada, E.~A., \& Lada, C.~J.\ 2001, ApJL, 553, L153 
\reference{} H{\o}g, E., Fabricius, C., Makarov, V.~V., et al.\ 2000, A\&A, 355, L27 
\reference{} Holland W. S., et al., 2017, MNRAS, 470, 3606
\reference{} J. D. Hunter, “Matplotlib: A 2D Graphics Environment,”Computing inScience \& Engineering, vol. 9, pp. 90–95, May 2007
\reference{} Jura, M., Ghez, A.~M., White, R.~J., et al.\ 1995, ApJ, 445, 451 
\reference{} Kalas, P.\ 2005, ApJL, 635, L169 
\reference{} Kennedy, G.~M., Bryden, G., Ardila, D., et al.\ 2018, MNRAS, 476, 4584 
\reference{} Kennedy, G.~M., Wyatt, M.~C., Sibthorpe, B., et al.\ 2012, MNRAS, 426, 2115
\reference{} Kennedy, G.~M., \& Kenyon, S.~J.\ 2008, ApJ, 682, 1264
\reference{} Kraus, A.~L., Shkolnik, E.~L., Allers, K.~N., \& Liu, M.~C.\ 2014, AJ, 147, 146
\reference{} Krivov, A.~V., et al. \ 2013, ApJ, 772, 32
\reference{} Krivov, A.~V., \& Booth, M.\ 2018, MNRAS, 479, 3300
\reference{} L{\'e}pine, S., \& Simon, M.\ 2009, AJ, 137, 3632
\reference{} Lestrade, J.-F., Wyatt, M.~C., Bertoldi, F., Dent, W.~R.~F., \& Menten, K.~M.\ 2006, A\&A, 460, 733
\reference{} Lestrade, J.-F., Matthews, B.~C., Sibthorpe, B., et al.\ 2012, A\&A, 548, A86
\reference{} Liu, M.~C., Matthews, B.~C., Williams, J.~P., \& Kalas, P.~G.\ 2004, ApJ, 608, 526
\reference{} Long, F., Pinilla, P., Herczeg, G. J. et al. 2018, ApJ 860, L17
\reference{} Luhman, K.~L., Stauffer, J.~R., \& Mamajek, E.~E.\ 2005, ApJL, 628, L69
\reference{} Luppe, P., Krivov, A.~V., Booth, M., et al.\ 2019, arXiv e-prints, arXiv:1910.13142
\reference{} MacGregor, M.~A., Wilner, D.~J., Chandler, C., et al.\ 2016, ApJ, 823, 79
\reference{} Marino, S., et al. 2019, arXiv:1909.09158
\reference{} Matr{\`a}, L., Marino, S., Kennedy, G.~M., et al.\ 2018, ApJ, 859, 72
\reference{} McMullin J. P., Waters B., Schiebel D., Young W., Golap K., 2007, Astronomical Society of the Pacific Conference Series Vol. 376, Astronomical Data Analysis Software and Systems XVI, p. 127
\reference{} Mamajek, E.~E., \& Bell, C.~P.~M.\ 2014, MNRAS, 445, 2169 
\reference{} Marshall, J.~P., Moro-Mart{\'\i}n, A., Eiroa, C., et al.\ 2014, A\&A, 565, A15
\reference{} Matthews, B.~C., Kennedy, G., Sibthorpe, B., et al.\ 2014, ApJ, 780, 97
\reference{} Matthews, B.~C., Kennedy, G., Sibthorpe, B., et al.\ 2015, ApJ, 811, 100
\reference{} McMullin, J. P., Waters, B., Schiebel, D., Young, W., \& Golap, K. 2007, Astronomical Data Analysis Software and Systems XVI (ASP Conf. Ser. 376), ed. R. A. Shaw, F. Hill, \& D. J. Bell (San Francisco, CA: ASP), 127
\reference{} Meshkat, T., Mawet, D., Bryan, M.~L., et al.\ 2017, AJ, 154, 245
\reference{} Meyer, M.~R., Calvet, N., \& Hillenbrand, L.~A.\ 1997, AJ, 114, 288
\reference{} Meyer, M.~R., Hillenbrand, L.~A., Backman, D., et al.\ 2006, PASP, 118, 1690 
\reference{} Morales, F.~Y., Bryden, G., Werner, M.~W., et al.\ 2016, ApJ, 831, 97
\reference{} Moro-Mart{\'i}n, A., Marshall, J.~P., Kennedy, G., et al.\ 2015, ApJ, 801, 143
\reference{} Ormel, C.~W., Liu, B., \& Schoonenberg, D.\ 2017, A\&A, 604, A1
\reference{} Pascucci, I., Laughlin, G., Gaudi, B.~S., et al.\ 2011, 16th Cambridge Workshop on Cool Stars, Stellar Systems, and the Sun, 448, 469 
\reference{} Pilbratt, G.~L., Riedinger, J.~R., Passvogel, T., et al.\ 2010, A\&A, 518, L1
\reference{} Plavchan, P., Jura, M., \& Lipscy, S.~J.\ 2005, ApJ, 631, 1161
\reference{} Plavchan, P., Werner, M. W., Chen, C. H., et al. 2009, ApJ, 698, 1068
\reference{} Phillips, N.~M., Greaves, J.~S., Dent, W.~R.~F., et al.\ 2010, MNRAS, 403, 1089
\reference{} Poglitsch, A., Waelkens, C., Geis, N., et al.\ 2010, A\&A, 518, L2
\reference{} Pollack, J. B., et al. 1996, Icarus, 124, 62 
\reference{} Raymond, S.~N., Armitage, P.~J., Moro-Mart{\'{\i}}n, A., et al.\ 2011, A\&A, 530, A62
\reference{} Riedel, A.~R., Finch, C.~T., Henry, T.~J., et al.\ 2014, AJ, 147, 85
\reference{} Riedel, A.~R., Murphy, S.~J., Henry, T.~J., et al.\ 2011, AJ, 142, 104 
\reference{} Riviere-Marichalar, P., Barrado, D., Montesinos, B., et al.\ 2014, A\&A, 565, A68 
\reference{} Roberge, A., Chen, C.~H., Millan-Gabet, R., et al.\ 2012, PASP, 124, 799
\reference{} Rodriguez, D.~R., Zuckerman, B., Faherty, J.~K., \& Vican, L.\ 2014, A\&A, 567, A20 
\reference{} S{\'a}nchez-Portal, M., Marston, A., Altieri, B., et al.\ 2014, Experimental Astronomy, 37, 453
\reference{} Schoonenberg, D., Liu, B., Ormel, C.~W., et al.\ 2019, A\&A, 627, A149
\reference{} Sibthorpe, B., Kennedy, G.~M., Wyatt, M.~C., et al.\ 2018, MNRAS, 475, 3046 
\reference{} Sibthorpe, B., Ivison, R.~J., Massey, R.~J., et al.\ 2013, MNRAS, 428, L6
\reference{} Silverberg, S.~M., Kuchner, M.~J., Wisniewski, J.~P., et al.\ 2016, ApJL, 830, L28 
\reference{} Smith, P.~S., Hines, D.~C., Low, F.~J., et al.\ 2006, ApJL, 644, L125 
\reference{} Su, K.~Y.~L., Rieke, G.~H., Stansberry, J.~A., et al.\ 2006, ApJ, 653, 675 
\reference{} Thureau, N.~D., Greaves, J.~S., Matthews, B.~C., et al.\ 2014, MNRAS, 445, 2558
\reference{} Torres, C. A. O., Quast, G. R., da Silva, L., et al. 2006, A\&A, 460, 695
\reference{} van Leeuwen, F.\ 2007, A\&A, 474, 653
\reference{} Wyatt, M.~C., Kennedy, G., Sibthorpe, B., et al.\ 2012, MNRAS, 424, 1206
\reference{} Zuckerman, B., Rhee, J.~H., Song, I., \& Bessell, M.~S.\ 2011, ApJ, 732, 61
\end{references}
\end{document}